# Theory of coagulation of metal nanodust clouds into micro-solids

*Peter V Pikhitsa*

e-mail: peterpikhitsa@gmail.com

**Abstract**

We describe briefly a possible coagulation mechanism for unipolar nanodust clouds of 3 nm single charged metal nanoparticles. A theoretical dependence of the critical cloud concentration vs the Hamaker's constant is derived and compared with available experimental data for Au, Ag, and Cu. Curious empty 1-3 um spheres, precipitated from the clouds in the result of coagulation, are discussed. The work may be interesting for nanoprinting and initial planetary formation mechanisms as establishing a bridge between nano and micro solids.

1. Introduction

Nanodust clouds (first introduced and named so in [1]) are mixture of unipolar charged metal nanoparticles of ~3-5 nm size, floating as a layer in an inert gas at atmospheric pressure over a same charge (and thus repelling) surface under a guiding electric field $E$ which delivers charged nanoparticles toward the surface and replenishes the floating layer with "fresh" nanoparticles. A close analogy is the charged dust layer floating over the lunar surface and observed by astronauts in Apollo mission [2] where the role of the field $E$ played the gravity field and the repulsion field came from the charged surface of the Moon. The nanodust clouds may appear in a nanoprinting procedure [3] where charged nanoparticles follow the guiding field toward the tips of the structures, growing on a conductive substrate, and cloud precipitation may lead to the surface contamination by uncontrolled cloud deposition and critical coagulation into micro (~2 um) metal spheres (super particles) [1].

The spheres are often either empty inside or having a lot of spherical pores, and sometimes round holes and/or other peculiar features on their surface (see SEM images in Suppl. to [1] and Fig. 1). Practically, the clouds can be dispersed and/or the micro spheres blown away by a strong inert gas flow, however, from a scientific point of view, the effect of coagulation into micro solids needs a deep investigation. There are reasons for that. First, the aggregation of nanoparticles into super particles may be a playground for solving a puzzle of initial planet formation [4]. Second, a modification of the nanoprinting process may put the coagulation under the electric field control to print useful and customized structures (other than just empty spheres) on metallic or even dielectric substrates where coagulation and the growth of solid structures can proceed even under the conditions of poor conductivity and thus poor dumping of charges from the growing structure.

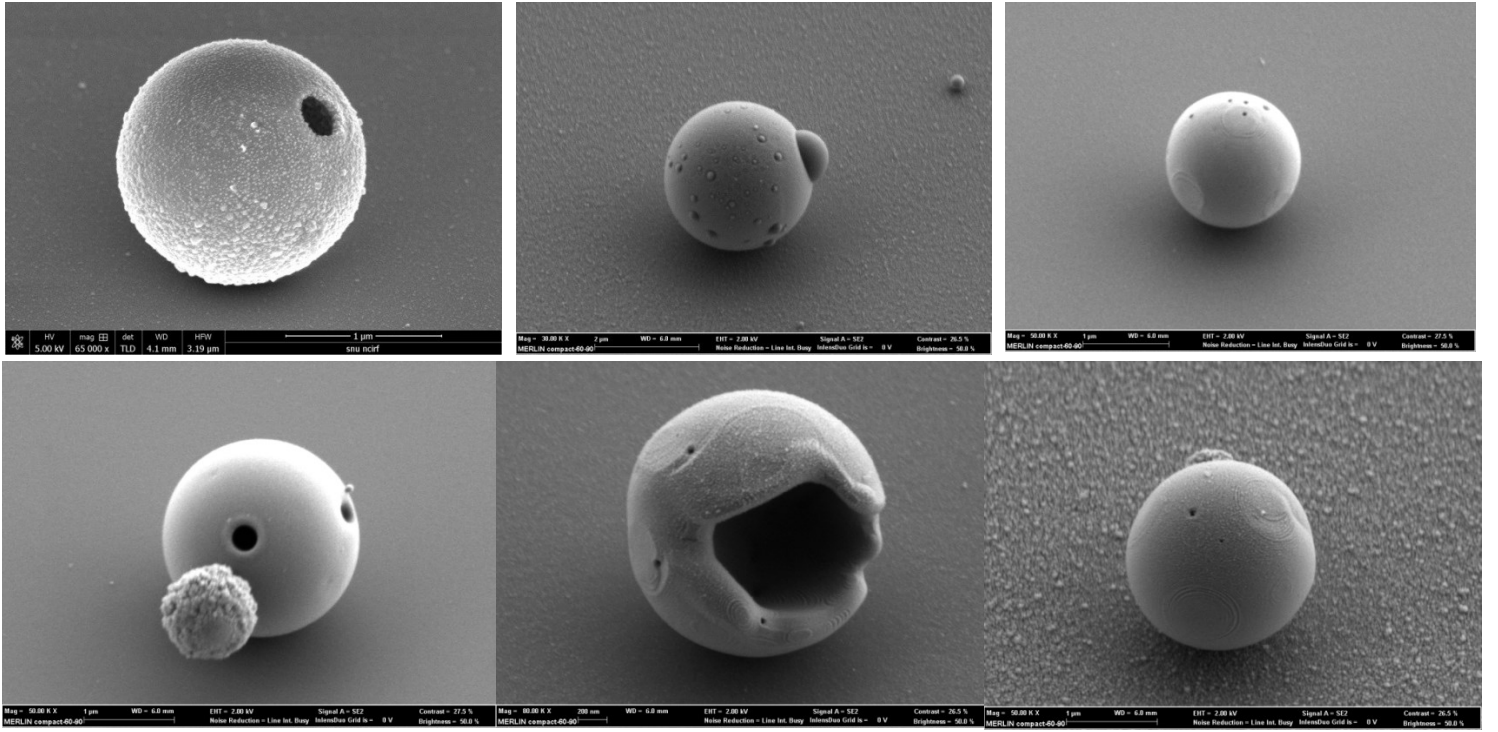

Fig. 1. SEM images of selected spherical super particles of Cu with peculiar features seen on the surface: holes, "swells", and "prints", indicating a complex fluid-like process of coagulation and solidification of nanodust.

Here we briefly present a theory of metal type dependent nanodust cloud coagulation which prescribes the mechanism of the clustering of unipolar nanoparticles to the van-der-Waals forces capable of attracting nanoparticles of the same charge to each other despite Coulomb repulsion, because the latter is Debye self-screened at large distances.

## 2. The Debye radius in unipolar plasma and DLVO theory of coagulation

Nanodust clouds were studied experimentally and theoretically in details in [1]. It was found that if a jet of nanoparticles wrote a stripe on a surface, then the width of the stripe always had a minimum at some strength of the applied electric field $E_{\text{dip}}$, called "dip field", which did not depend on the experimental parameters except on the metal type of nanoparticles. The phenomenon of coagulation can explain the curious universality of the "dip field" because $E_{\text{dip}}$ was theoretically connected in [1] with the concentration of the dust by the equation:

(1) $E_{\text{dip}} = 2\pi \sqrt{\dfrac{k_\text{B} T n}{\varepsilon_\text{r} \varepsilon_0}}$ ,

where $\varepsilon_0$ is the permittivity of free space, $\varepsilon_r \approx 1$ is the relative gas permittivity, $k_B$ is the Boltzmann constant, $T$ is the temperature in Kelvins, and $n$ is the steady state concentration of the singly charged nanoparticles (charge $e$) in the nanodust cloud at a given gas flow velocity.

Physically, $E_{dip}$ is the value of the applied field strength that provides the narrowest focusing of coming nanoparticles onto the substrate surface. According to Eq. (1), the steady state concentration $n$ should also be constant and depend only on the type of metal. One can assume that such constancy is due to the self-sustained coagulation and the following precipitation that starts when the cloud reaches some critical coagulation concentration [5], and keeps at the same steady state critical value during nanoprinting because of coagulation of extra nanoparticles into metal spheres of 1-3 um that fall out of the cloud, not unlike of what happens to rain clouds. The whole process looks like a self-organized criticality.

The Coulomb interaction in unipolar dust plasma in the nanodust cloud is screened on the Debye radius [1]:

(2) $r_D = \sqrt{\dfrac{\varepsilon_r \varepsilon_0 k_B T}{e^2 n}}$ .

The existence of the Debye radius allows one to apply the DLVO criterium (see page 329 in [5])

(3) $k^3/\rho_\infty = 768\pi k_B T \gamma^2 \mathrm{e}^{-1}/A$

for coagulation of spherical particles to the case of the nanodust cloud:

(4) $A = 768\pi k_B T \gamma^2 \mathrm{e}^{-1} n^{-1/2} \left(\dfrac{\varepsilon_r \varepsilon_0 k_B T}{e^2}\right)^{3/2}$,

where in Eq. (3) we replaced the inverse Debye radius $k = 1/r_D$ with Eq. (2) and concentration $\rho_\infty$ with $n$. In Eqs. (3) and (4) $A$ is the Hamaker constant for the metal of the nanoparticles, e is the Euler's number, and $\gamma = \tanh\left(\dfrac{e\psi_0}{4k_B T}\right) \approx \dfrac{e\psi_0}{4k_B T}$ for small surface potential $\psi_0 \ll 100 mV$ at room temperature. Now one can rewrite Eq. (4) in terms of experimentally observable $E_{dip}$ from Eq. (1) to obtain:

(5) $A = 768\pi \mathrm{e}^{-1} \dfrac{\psi_0^2}{16e} \dfrac{2\pi}{E_{dip}} \varepsilon_r \varepsilon_0 k_B T = 24\pi \mathrm{e}^{-1} \dfrac{(e\psi_0)^2}{(eE_{dip}\lambda_B)}$,

where we introduce the Bjerrum length $\lambda_B = e^2/(4\pi \varepsilon_r \varepsilon_0 k_B T)$ just to show explicitly that Eq. (5) is dimensionally correct, because the square of energy on the rhs is divided by the energy to give the Hamaker constant on the lhs. Note, that the only fitting parameter in Eq. (5) is the surface potential $\psi_0$.

## 3. Comparison with known experimental data and some predictions

We take the experimental values [1] for $|E_{dip}| = 0.4\frac{kV}{cm}; 0.7\frac{kV}{cm}; 0.8\frac{kV}{cm}$ for three metals Au, Ag, Cu, respectively. The Hamaker constants for these metals are $A = 402\ zJ; 368\ zJ; 338\ zJ$, respectively [6]. The best fit is given in Fig. 2 with the value for the surface potential of $\psi_0 = 15\ mV$. Note, that although Pd was the case in [1] as well, we currently eliminate it from consideration because Pd is known to produce dimers of nanoparticles in the process of the spark discharge generation that may modify the nanodust cloud and the coagulation process.

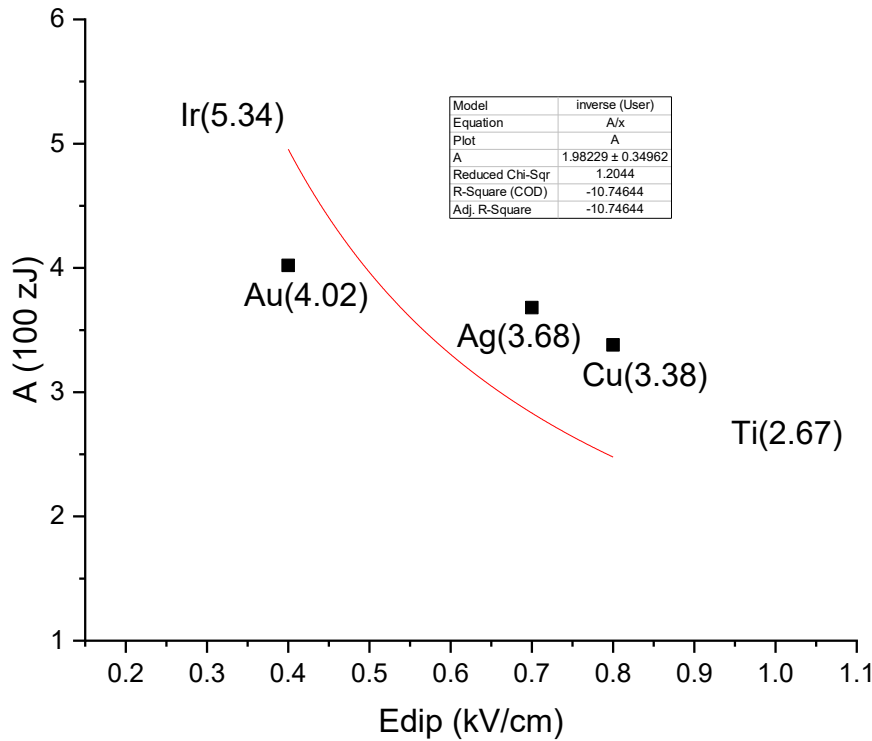

Fig. 2. Hamaker's constant vs the dip field (black squares). The red curve is the best fit with Eq. (5). The places for Ir (the largest Hamaker's constant) and Ti (the smallest Hamaker's constant) are left empty and there is a need for experimental values for Edip for these metals and other metals.

The inverse dependence of Eq. (5) can be interpreted with the fact that smaller van-der-Waals interaction needs shorter distance between nanoparticles for coagulation, therefore larger concentration and larger $E_{dip}$.

It would be interesting to investigate mechanical properties of super particles resulting from coagulation of nanodust clouds. From their shapes and forms it is clear that during formation many violent processes took place: inert gas trapped with individual nanoparticles along with the surface charge produced swells on the surface (as seen in Fig. 1) with gas bubbles "frozen" inside super particles [1]. Sometimes the swells collapsed and left holes, sometimes they just deflated and left flat round printed circles of the surface (Fig. 1). Cutting perfect 2 um Cu spheres with Focused Ion Beam (FIB) demonstrated a perfect shell of 200 nm [1] which closely resemble soap bubbles, implying fluid-like or colloid-like behavior of coagulating material before it is "frozen". Moreover, an SEM image of a large spherical void inside an FIB cut Au super particle in Fig. 3 (left) reveals a hexagon structure (recall the famous Saturn polar hexagon or hexagonal basalt columnar joining in volcanic rocks) on the surface of the void. Such a regular structure may be a footprint of complicated solidification patterns in coagulating colloid. Indeed, the appearance of a pentagon in a void of an Ag super particle in Fig. 3 (right) is a natural variation (a defect) in hexagonal basalt columnar joining.

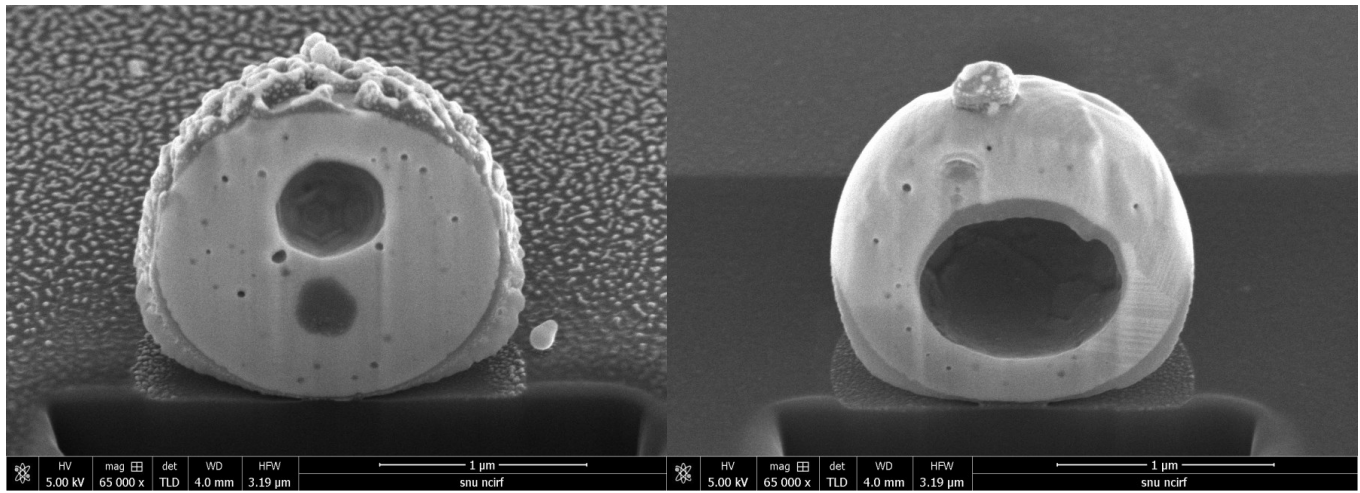

Fig. 3. A hexagon structure seen in the void inside an FIB cut Au super particle (left) and a pentagon in the void of an Ag super particle (right).

4. Conclusions

We demonstrated that nanodust clouds may have a complex colloid-like behavior and be prone to coagulation into microscopic spheres. Such coagulation explains a constant critical concentration of nanoparticles in the cloud and its material dependence, observed experimentally before. Controlling nanodust colloid with electric fields may be useful for metal nanoprinting on poor conducting and even dielectric substrates.